# A New Bayesian Optimization Algorithm for Complex High-Dimensional Disease Epidemic Systems and Related Computational Studies

Yuyang Chen, Kaiming Bi, Chih-Hang J. Wu, David Ben-Arieh, Ashesh Sinha
Kansas State University, USA
cyy@ksu.edu

**Abstract**
This paper presents an Improved Bayesian Optimization (IBO) algorithm to solve complex high-dimensional epidemic models' optimal control solution. Evaluating total objective function value for disease control models with hundreds of thousands of control time-periods is high computational cost. In this paper, we improve the conventional Bayesian Optimization (BO) approach from two parts. The existing BO methods optimize the minimizer step for once time during each acquisition function update process. To find a better solution for each acquisition function update, we do more local minimization steps to tune the algorithm. When the model is high dimensions, and the objective function is complicated, only some update iterations of the acquisition function may not find the global optimal solution. The IBO algorithm adds a series of Adam-based steps at the final stage of the algorithm to increase the solution's accuracy. Comparative simulation experiments using different kernel functions and acquisition functions have shown that the Improved Bayesian Optimization algorithm is effective and suitable for handing large-scale and complex epidemic models under study. The IBO algorithm is then compared with four other global optimization algorithms on three well-known synthetic test functions. The effectiveness and robustness of the IBO algorithm are also demonstrated through some simulation experiments to compare with the Particle Swarm Optimization algorithm and Random Search algorithm. With its reliable convergence behaviors and straightforward implementation, the IBO algorithm has a great potential to solve other complex optimal control problems with high dimensionality.

*Keywords:* Bayesian improvement, complex system, high-dimension, optimal control

## 1. Introduction

Many complex and high-dimensional optimization problems require to find the extremum of a system that includes multiple local optima using as few tries as possible. Recently developing more effective and efficient global optimization approaches get more attention [1]. The important thing is that global optimization algorithms should have the capacity to capture a better solution at each optimization process to seek out the global optimum soon. The optimization processes need to solve complex nonlinear systems with large numbers of variables or constraints and frequently costly to evaluate their objective functions [2-4]. These previously mentioned challenges inspired the necessity to develop more efficient and robust global optimization algorithms and streamline the implementation processes to get the best performance and solve the complicated systems' optimal solutions.

In the past few decades, optimization problems raised in various real-world applications have been studied. Many state-of-the-art global optimization methods have been proposed in the literature, such as the Cutting plane method [5], Branch and bound methods [6], Monte-Carlo methods [7], Genetic algorithms [8], Simulated annealing [9], Particle swarm optimization [10]. Several improved algorithms have been developed based on those global optimization algorithms to handle combinatorial or global optimization problems raised in real-world applications. An extended cutting plane method is introduced and applied to solve convex mixed-integer nonlinear programming problems [11]. A bi-objective branch-and-bound method was proposed for solving a subclass of multi-objective mixed integer programming problems with two objectives allowed and binary variables [12]. In [13], the Monte Carlo algorithms with Markov Chains are used to study a residential distribution circuit and determine a pattern to energy demand. A variety of improvements have been proposed to improve the Particle Swarm Optimization (PSO) algorithm or address its shortcomings by combining genetic algorithms. For example, a genetic learning Particle Swarm Optimization is proposed to use the genetic evolution to breed exemplars to PSO algorithm, then guidance the exemplars by the historical search experience of particles [14]. A new approach is developed by combining the Genetic algorithm and PSO algorithm to find the most discriminative features using feature selection [15]. These algorithms help them to reach a good performance for the lower-dimensional or straightforward systems. However, for high dimensional and complex dynamic systems, they also suffer from relatively long running time or usually trap at a local minimum before reaching the real optimal solution.

The optimal control problem is a typical high-dimensional case in the healthcare field, such as optimal control of epidemic problem [16], optimal control of COVID-19 [17], optimal control of sepsis treatment [18], optimal control of HIV [19], etc. In general, a control variable in those problems contains many time-periods, even up to hundreds of thousands of time-periods for one control strategy. Each time-period is considered as a dimension to the model. When there are many state variables in the model, it is necessary to calculate the value of state variables at each time-period and sum up each period's cost to evaluate the overall cost of a control strategy. It could be too time-consuming to evaluate the overall cost, even for a single control strategy. Optimal control problems are often more complex and hard to solve when both objective functions and constraints are nonlinear and possible non-convex—for example, the optimal control of a disease dynamic epidemic model.

An infectious disease epidemic frequently lasts for hundreds of days or even a couple of years. The controls or interventions for these disease models could be as high-dimensional since the interventions would be carried out in each time-period. During the epidemic, the health organizations or agencies may take a series of measures to decide the level of controls for each local outbreak, e.g., vaccination, quarantine, disinfection, or regional closures. All these control measures could be associated with certain financial costs, directly or indirectly. If health organizations or agencies do not control the epidemic, it may also cause inevitable economic consequences, such as workforce losses due to outbreaks, increased community healthcare costs, local business downturns, and declined related travels. Thus, an optimal control problem's goal is to balance the cost of control and the cost of null control, and the associated objective function is usually not convex or unimodular. It would be useful if one can solve the optimal control using accumulated historical experiences and benefit from the prior solution statistically. The above challenges hinted us to use Bayesian Optimization and apply it to solve epidemic models' optimal control problems.

Bayesian optimization (BO) utilizes an acquisition function to approximate the complex objective function. It has become a prevalent method for global optimization. Standard BO is a novel framework dealing with both exploration and exploitation and determines the proper domain spaces to sample during the search process. Related computational studies from existing literature have shown great success for some problems where the objective function evaluations are very costly, or the objective function lacks an analytical expression [1]. Moreover, recently BO algorithm is further extended to achieves better optimization results. In 2017, Bayesian optimization is used as approach for modeling construction of data-driven stochastic evolution model [20]. A weighted expected improvement-based BO algorithm is proposed to handle multi-objective optimization problems [21]. In [22], a new BO algorithm leveraging gradients in hyperparameter tuning is proposed to reduce the number of objective function evaluations in low dimensional problems. For multi-response surface optimization problems, a new BO approach incorporating both expected loss and its variances into a Bayesian modeling uniform framework was also proposed [23]. This proposed approach considers the uncertainty of model parameters and measures the reliability of an acceptable optimization result. A BO algorithm with an elastic Gaussian process is also introduced and tested in the optimization problem with less than a hundred dimensions [24]; this algorithm enables local gradient-dependent algorithms to move through the flat terrain. For COVID-19, researchers combined deep learning and Bayesian optimization to predict the COVID-19 time-series data [25]. However, BO applied in these studies mentioned above is limited for solving lower dimensional systems, or some of them require extended evaluation or calculation time to reach their best performance. According to the concept and implementation process of BO, some aspects significantly impact its performance and solution accuracy, such as acquisition function, kernel function, sampling method, and tuning in the implementation process.

Motivated by the challenges and existing knowledge of BO algorithms discussed above, we proposed an improved version of BO and utilized the improved algorithm to solve an SEIR epidemic model's optimal controls. The contributions of this paper are as follows:
(1) We propose an improved version BO algorithm. The improved algorithm is demonstrated through some computational experiments that it can solve the global optimal solution for both low-dimensional optimization problems and complex high-dimensional optimal control models within the limited number of iterations.
(2) We do several suboptimal steps in the IBO algorithm to optimize the acquisition function. In addition, we add a series of Adam-based steps as a local search after acquisition function optimization, which aims to increase the accuracy of the final optimal solution. Those improvements are proved to be effective by comparing with simplicial homology global optimization, dual annealing optimization, differential evolution, basin-hopping algorithms from Scipy Optimize library [26] on three well-known synthetic single-objective test functions three well-known synthetic single-objective test functions, and also comparing with PSO algorithm, Random Search algorithm, standard BO algorithm on the high-dimensional epidemic control model.
(3) The proposed IBO algorithm is demonstrated that performs robustly with various initial control strategies and it may perform slight different optimization results when it selects different kernel functions and acquisition functions.

The remainder of this paper is organized as follows. Section II formulates the high-dimensional optimal control problem. In Section III, the state-of-the-art related works are summarized and commented. Section IV presents the IBO algorithm framework in detail with several implementation insights. Then, the computation studies and their results are presented to demonstrate the effectiveness and efficiency of the IBO algorithm in Section V. Conclusions and potential future studies are summarized in this final section.

## 2. Problem formulation

Consider the system control period $[0, t_f]$, $t_f$ is the final time. The general high-dimensional optimal control problem can be formulated as follows:

$$\min \quad V(x, u) \quad (1)$$
$$s.t. \quad \frac{dx}{dt} = g(x, u) \quad (2)$$

where $x \in R^d$ represents $d$ state variables, time $t \in [0, t_f]$, control $u = \{u(0), \dots, u(t_f)\}$, $u(t) \in [u_l, u_u]$ represents the level/degree of the control measure (such as vaccination, quarantine, disinfection, or regional closures) at time $t$, $u_l$ and $u_u$ are the lower and upper bound of control strategies, respectively. $V(x, u)$ represents the overall cost of the system during $[0, t_f]$. $V(x, u)$ in this paper is considered as a high-dimensional non-convex function. For simplification, we only consider the minimization problem in this paper. A maximization problem can be easily converted to a minimization problem by setting the negative of the objective function $V(x, u)$.

In this paper, we study the optimal control problem for an SEIR control model during $[0, t_f]$ according to an SEIR epidemic model in [27]. Our main goal is to control the infected population through possible disease intervention strategies, such as vaccination, quarantines, or safeguard procedures. The cost function $V(x, u)$ describes the total cost due to infection and control strategy. The parameters $C_1$ and $C_2$ represent the cost of null control and the cost of control per individual in each time-period, respectively. Thus, the problem can be formulated as follows:

$$\text{Min } V = \int_0^{t_f} C_1 I(t) + C_2 \left| 0.3 \sin(10u(t)) + \sin(13u(t)) + 0.9 \sin(42u(t)) + 0.2 \sin(12u(t)) + u^2(t) \right| \quad (3)$$

$$s.t. \quad \frac{dS(t)}{dt} = \tau - \beta S(t)I(t) - \tau S(t) \quad (4)$$
$$\frac{dE(t)}{dt} = \beta S(t)I(t) - (\tau + \alpha)E(t) \quad (5)$$
$$\frac{dI(t)}{dt} = \alpha E(t) - (\tau + \gamma)I(t) - u(t)I(t) \quad (6)$$
$$\frac{dR(t)}{dt} = I(t) - \tau R(t) + u(t)I(t) \quad (7)$$
$$S(t) + E(t) + I(t) + R(t) = 1 \quad (8)$$

where $S(t), E(t), I(t), R(t)$ represent the fraction of susceptible, exposed, infectious and recovery population at time $t$, respectively. Assume that the natural birth rate and natural death rate are identical as $\tau$. Parameter $\beta$ is the contact rate. Population transfers from $E$ to $I$ with probability $\alpha$ in a small interval $dt$. Population transfers from $I$ to $R$ with probability $\gamma$ in a small interval $dt$.

To provide an intuitive display of the control model studied in this paper, we show a 3-dimensional plotting of our system for two time-periods in Fig. 1. We can see that only the 3D model is complex and with multiple local optima. However, our studied model is with $t_f$ dimensions where practically $t_f$ could equal to several hundred time-periods.

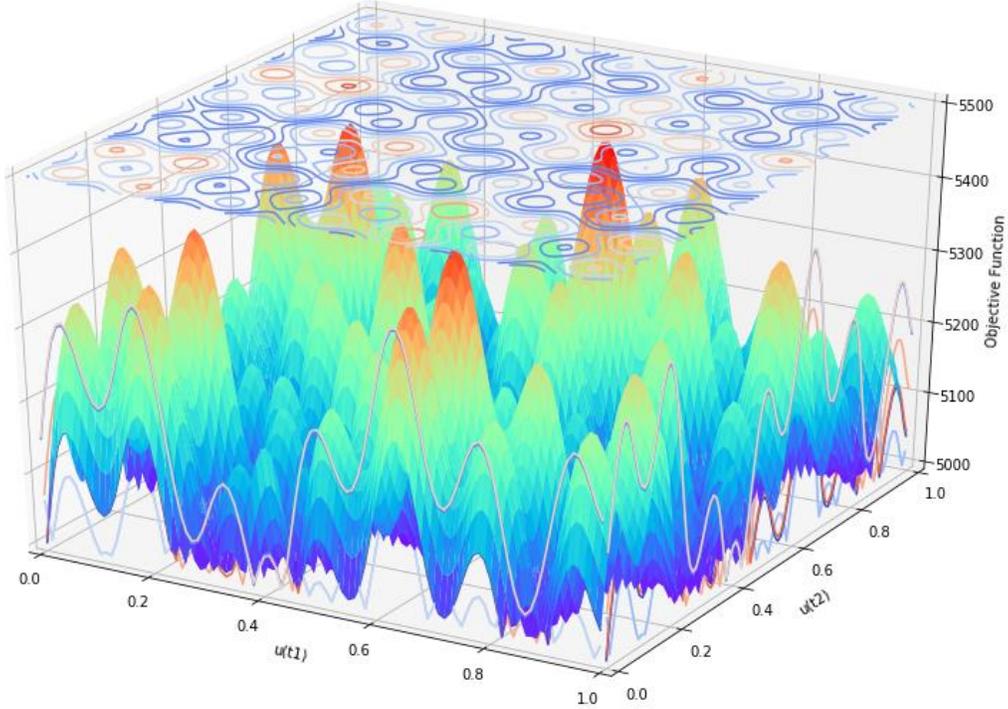

Fig. 1. 3-dimensional plotting of SEIR control model

## 3. The related work

This section reviewed the essential concepts and purposes for the IBO algorithm and related works in this area. The related results of choosing the given high-dimensional control system are also discussed in the section. The standard BO algorithm framework is thoroughly reviewed in the article [28]; there are two essential parts for BO: the probabilistic surrogate model of objective function and acquisition function. The probabilistic surrogate model evaluates the model uncertainty based on the observed sampling data, and the acquisition function mainly balances the exploration and exploitation during the optimization process [27]. We briefly summarize the general procedure as follows: (1) Initial a start sampling point; (2) Construct the probabilistic surrogate model. Generally, the posterior model is considered as the probabilistic surrogate model of objective function $V(x, u)$ for Bayesian optimization; (3) Optimize the acquisition function to select a next sampling point; (4) Calculate the corresponding objective function value $V(x, u)$ for this sampling point; (5) Update the probabilistic surrogate model by adding the new sampling point and corresponding objective function value; (6) Repeat (3) – (5) for some iterations and return the best objective function value and corresponding sampling point.

To improve the standard BO algorithm, we presented various studies with different structures of the surrogate models, the kernel functions, the acquisition function, and the sampling strategies, and then test most of them in the give high-dimensional optimal control models.

### 3.1 Gaussian Processes

There are many methods to construct the surrogate model, such as neural networks, support vector machines, random forests, and the Gaussian Process (GP). When a function follows a GP, then the likelihood is Gaussian, and its posterior also is a GP. Since the property and flexibility of the GP, it becomes a common and popular choice of surrogate model for BO.

The GP is a probability distribution over function. Assume the original objective function $V(x, u)$ follows a GP, consider $V(x^i, u^i)$ as $V(u^i)$ when the control strategy is $u^i = \{u^i(0), \ldots, u^i(t_f)\}$. Then

$$V(u^i) \sim \mathcal{GP}(m(u^i), k(u^i, u^{i'})) \tag{9}$$

where $m(u^i)$ is mean function and $k(u^i, u^{i'})$ is covariance function [29], where $u^i$ and $u^{i'}$ represent two different control strategies. The mean function is usually defined as a linear function or directly defined as zero [30]. The covariance function is also named kernel function.

Any finite number of the objective function values, $V(u^i)$, follow multivariate Gaussian distribution [31]. Let $V = [V(u^0), ..., V(u^i)]^T$, then $V$ is Gaussian distributed with mean $M = [m(u^0), ..., m(u^i)]^T$ and covariance matrix $K$ as below:

$$K = \begin{bmatrix} k(u^0, u^0) & \cdots & k(u^1, u^i) \\ \vdots & \ddots & \vdots \\ k(u^i, u^0) & \cdots & k(u^i, u^i) \end{bmatrix} \quad (10)$$

For any new sampling point $u^j$ and corresponding $V(u^j)$, let

$$V' = \begin{bmatrix} V \\ V(u^j) \end{bmatrix}, M' = \begin{bmatrix} M \\ m(u^j) \end{bmatrix}, \Sigma = \begin{bmatrix} K & K'^T \\ K' & K'' \end{bmatrix} \quad (11)$$

where $K' = [k(u^j, u^1), k(u^j, u^2), ..., k(u^j, u^i)]$, $K'' = k(u^j, u^j)$.

Then the posterior distribution of $V(u^j)$ for any new sampling point $u^j$ based on all known data $V$ will Gaussian distributed with mean $\mu(V(u^j)|V)$ and variance $\sigma(V(u^j)|V)$, which can be written as:

$$V(u^j)|V \sim \mathcal{GP}(\mu(V(u^j)|V), \sigma(V(u^j)|V)) \quad (12)$$

where the posterior mean and the variance can be derived as:

$$\mu(V(u^j)|V) = m(u^j) + K'K^{-1}(V - M) \quad (13)$$
$$\sigma(V(u^j)|V) = K'' - K'K^{-1}K'^T \quad (14)$$

### 3.2 Choices of Kernel function

The kernel function is an essential part of the Gaussian Process modeling, and it performs differently for different modeling applications. There are several kernel function choices that can be used in BO algorithm, such as Matern32, Matern52, Radial Basis Function (RBF), Exponential, Linear, Brownian, Periodic, Polynomial, Warping, Coregionalize, RationalQuadrati (RQ). In most existing literatures about the BO algorihtm, Matern32, Matern52 and Radial Basis Function (RBF) are the three more popular choices [32]. The impacts of different kernel function choices on the global performance of the IBO algorithm will be detailed discuss in the simulation part.

### 3.3 Acquisition functions

The acquisition function takes the probabilistic mean and variance at each sampling point on the objective function. It can be used to evaluates how desirable the next sampling position would be. An adequately designed acquisition function for the BO model should represent a trade-off between exploration and exploitation. The exploration suggests that objective function values could be highly uncertain. On the other hand, exploitation implies that the next sampling point could have a lower objective function value for the minimization problem [33]. Several popular approaches can be applied as the acquisition function including, Lower Confidence Bound (LCB), Probability of Improvement (PI), Expected Improvement (EI), Entropy Search, and Thompson sampling. The impact of different acquisition functions on the IBO algorithm is analyzed in the simulation part via some computational experiments.

### 3.4 Candidate sampling strategies

Using an acquisition function to estimate the complex objective function values may still encounter difficult issues when dealing with high-dimensional optimal control models. It is impossible to traverse the entire feasible solution space for a high-dimensional model in a reasonable amount of time. An efficient and effective sampling approach is the key to attack complex, large-scale global optimization problems. Other research works have suggested many existing algorithms as sampling strategies, such as Random Search, Sequential Search, PSO algorithm, Simulated Annealing, and Tabu Search. Also, other ideas related to distance or distribution can be applied.

This paper leverages the idea of distribution to select the candidates for the next sampling point. Let $\mathcal{D}$ be s set of known sampling points. $(u^{last}, V(u^{last}))$ be the last data of $\mathcal{D}$, which is also the optimal sampling point selected from the last optimization iteration. Consider $(u^{last}, V(u^{last}))$ as one of the candidates for the next sampling point. Specify a uniform distribution for each dimension of $(u^{last}, V(u^{last}))$, and randomly sample $k$ values for each dimension from this distribution, and then compose $k$ sampling candidates. For each candidate, consider the acquisition function as the loss function. *Autograd.backward* function in Python is used to compute the accumulated gradients for the loss function. After that, apply the result information returned from *Autograd.backward* function to update the step for $k$ times when we do the gradient descent for the acquisition function. Eventually, select $l$ $(0 < l < k)$ candidates with the lowest acquisition function values as the next sampling points and add them into the database for the update of the GP model.

## 4. IBO algorithm framework for high-dimensional optimal control models

Frequently, it could be time-consuming to repeatedly evaluate the objective function values for complex and high-dimensional models. Instead, one can use its information to establish a simpler approximation function – acquisition function – during the optimization process to serve as a springboard in conserving computational efforts.

Most BO algorithms in the existing literature perform a single optimization step for the acquisition function when searching for the next sampling point. Due to the complexity and high dimensions of the objective function, single-step optimizing the acquisition function would yield unreliable sampling points that are not worthwhile to explore and ultimately result in a longer running time. The IBO algorithm attempts to perform several optimizations for each sampling candidate based on the acquisition function. This improvement of the IBO algorithm can efficiently pick out the better sampling point when the model is high-dimensional.

Besides, most successful BO approaches only handled the iterations to train the GP model and optimize the acquisition function once for each sampling candidate Then, among few candidate sampling points, by optimizing the acquisition function to pick the optimal smaple candidate with the lowest acquisition function value. However, for the high dimensional complex model, one cannot explore the entire solution space to find the optima for the acquisition function in a reasonable amount of time. Also, the solution we select on acquisition function optimization part is in terms of the acquisition function values rather than the original objective function values. Thus, to increase the accuracy of final optimal solution, a series of Adam-based step is added as a local search process after the acquisition function optimization. This local search process will use the original objective function instead of the acquisition function.

The IBO algorithm framework for the high-dimensional optimal control problem can be summarized in detail as follows:

**Algorithm 1** The Improved Bayesian Optimization Algorithm

1: Randomly initial some control strategy inputs with multi-dimensions $(u^0, u^1, ..., u^j)$
2: Compute the state variables through the control inputs and the given high-dimensional control model
3: Calculate the corresponding objective function values for each control strategy input and same them to compose
  a dataset $\mathcal{D} = \{(u^0, V(u^0)), (u^1, V(u^1)), ..., (u^j, V(u^j))\}$
4: Train the Gaussian process model by using dataset $\mathcal{D}$
5: **for** i = 1, 2, ..., $m$ **do**
6:   Select control $u$ from the last data in dataset $\mathcal{D}$ to be one of the new candidates for the next sampling points
7:   Randomly generate $k$ sampling candidates from the uniform distribution
8:   **for** each candidate **do**
9:     **for** p = 1, 2, ..., $n$ **do**
10:       Calculate the acquisition function value
11:       Use *autograd.backward* to obtain the derivate information of acquisition function and then update
          the step size of optimizer for acquisition function
12:     **end for**
13:   **end for**
14:   Find $l$ candidates with the lowest acquisition function values to be the next sampling points
15:   Add these $l$ candidates and the corresponding objective function values to the dataset $\mathcal{D}$
16:   Update the Gaussian process model
17: **end for**
18: **Obtain** a best control $u$ with the lowest acquisition function value from all sampling points during iterations
19: **Repeat**
20:   Use *autograd.backward* to obtain the derivate information of the objective function, and update the step
      size to do the local optimization for the objective function
21:   **Until** the objective function $V$ converges
22: **Return** the global optimal control solution after while The IBO algorithm was implemented in Python 3.7 using Pytorch and Pyro libraries. The codes were executed on a Personal Computer with Intel i5 Center Process Unit and 32 GB of Random Access Memory for more than a hundred times to find the more suitable values of parameter $m$ – main loops, the number of candidates, $k$, and parameter $n$ – number of local minimization steps. For the given high-dimensional control system, our experiment results show that the IBO algorithm can almost find the global optimal solution when $m$ is about 15, the number of candidates is around 5, $n$ is set to 5.

## 5. Simulation

In this section, some simulation experiments are conducted to evaluate the performance of the IBO algorithm. Part A implements the IBO algorithm and some other optimization algorithms on three common test functions. The purpose of Part A is to prove that the IBO algorithm is effective for the low-dimensional global optimization problem, and it can accurately solve the global optimal solution. Part B is simulations to demonstrate the convergence of the IBO algorithm in high-dimensional control system. The simulations in Part C are conducted to study the IBO algorithm's general performance by comparing it to the Random Search algorithm and PSO algorithm. All simulation experiments are conducted on Python version 3.7 with Intel Core i5 CPUs and 32G memory. The kernel function selected in the following simulation experiments is Matern52, and the lower confidence bound function is defined as the acquisition function.

### 5.1 Impact of different kernel functions on the global performance of the IBO algorithm

To study the impacts of different kernel function choices on the rglobal performance of the IBO algorithm, several kernels types that were tested in this section. For each kernel function choice, we repeatedly implemented tests on the given high-dimensional control model 20 runs with the same initial conditions and model parameters values, and summarized the mean of all results for each kernel function. All simulation experiments were implemented and tested in Python version 3.7 programming language. Fig. 2 shows the average objective function values for different kernels choices. Our computational experiments have shown that the RBF, Warping, and Matern52 kernels produce better objective function values than others for the given control model. Fig. 3 shows the average running times for different GP kernels. Brownian, RBF, Linear, and Coregionalize are more efficient than the other kernels. Consider those simulation results, and we can see that RBF and Matern52 reach a better all-around performance than other kernel choices. Both RBF and Matern52 kernels are stationary kernels. RBF kernel is also known as the "squared exponential" kernel. Matern kernel is a generalization of the RBF kernel that contains an additional parameter that can control the resulting function's smoothness. Therefore, they are two popular kernel options for the GP.

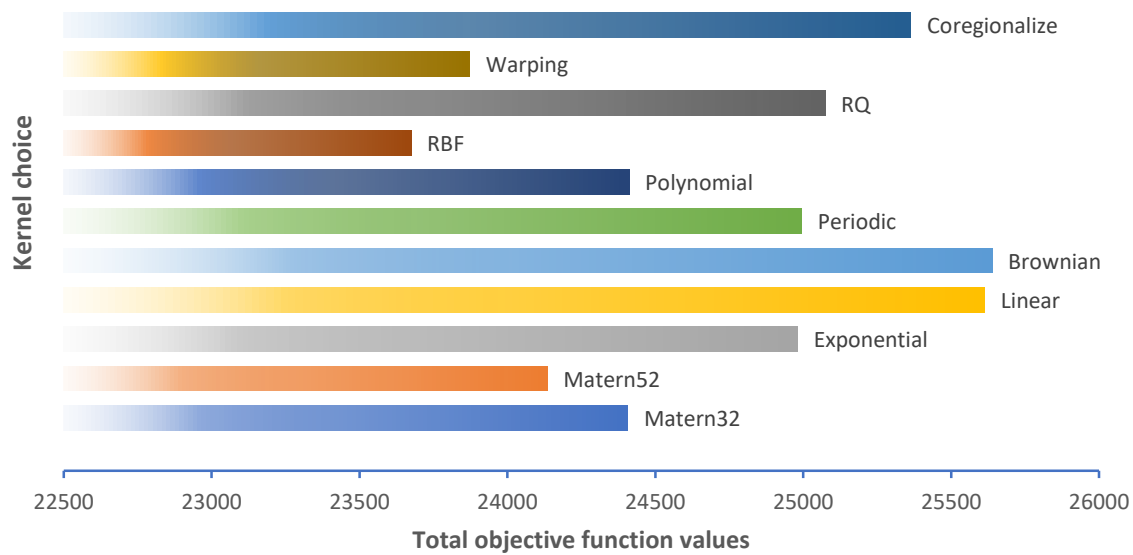

Fig. 2. The mean of total objective function values for different kernels choices

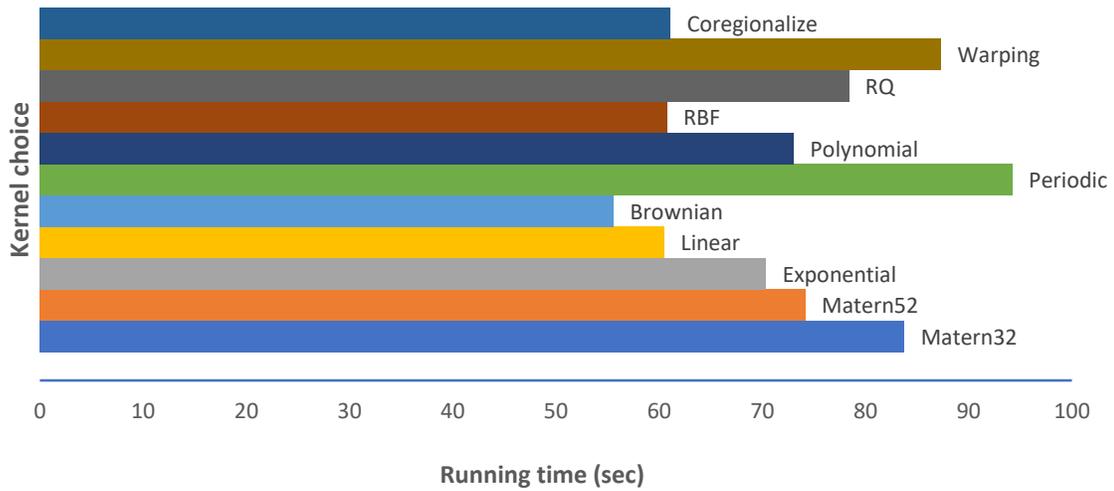

Fig. 3. The running time for different kernels choices

### 5.2 Impact of different acquisition functions on the global performance of the IBO algorithm

To test the influences of the acquisition functions on the IBO algorithm, some simulation experiments were carried out using the three most popular acquisition functions: LCB, PI, and EI. For each acquisition function, 20 replications of simulation experiments are conducted by Python and summarized as the average performance of more than 50 iterations. The results of three different acquisition function choices after 25 iterations tend to be constant and almost coincide. Thus, we will only show up the results of 26 iterations. Fig. 4 shows the results for three different acquisition function choices within 26 iterations. Although these three acquisition functions perform differently before the first iterations, they reach similar optimization results eventually after about 20 iterations. Thus, For the IBO algorithm, three different acquisition function choices do not have impacts on the final results. In addition, all three results have a sudden downward trend at about the 16[th] iteration. Before they decrease sharply, the lines represent the results during the acquisition function optimization iterations. From about the 16[th] iterations, the smooth descent curves represent the results when the IBO algorithm does the local search. It shows that adding the local search after the acquisition function optimization part is important and necessary for the accuracy of the final solution.

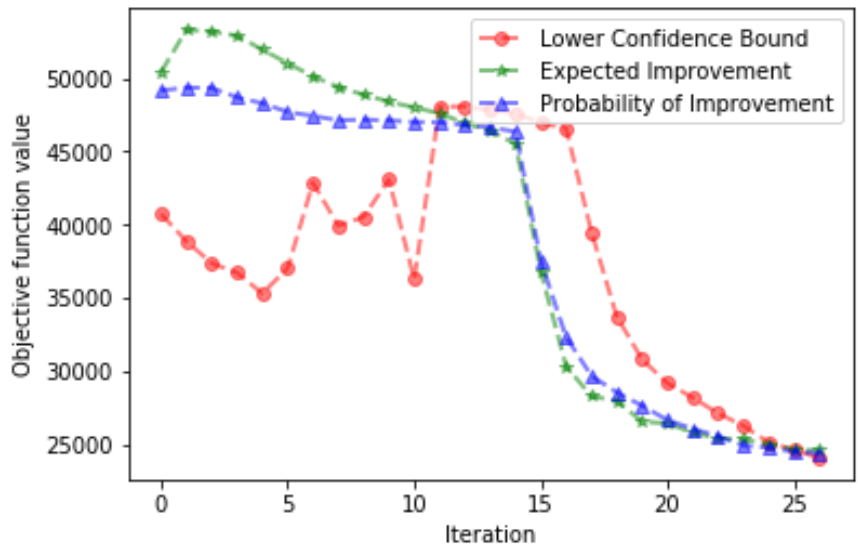

Fig. 4. The convergence performance for different acquisition functions choices

Figure 5 provides a 3-dimensional display of how does IBO algorithm chooses the sampling point position using an acquisition function. The green surface represents the original objective function, while the grey surfaces represent the surface that passes through the sampling points generated from the acquisition function.

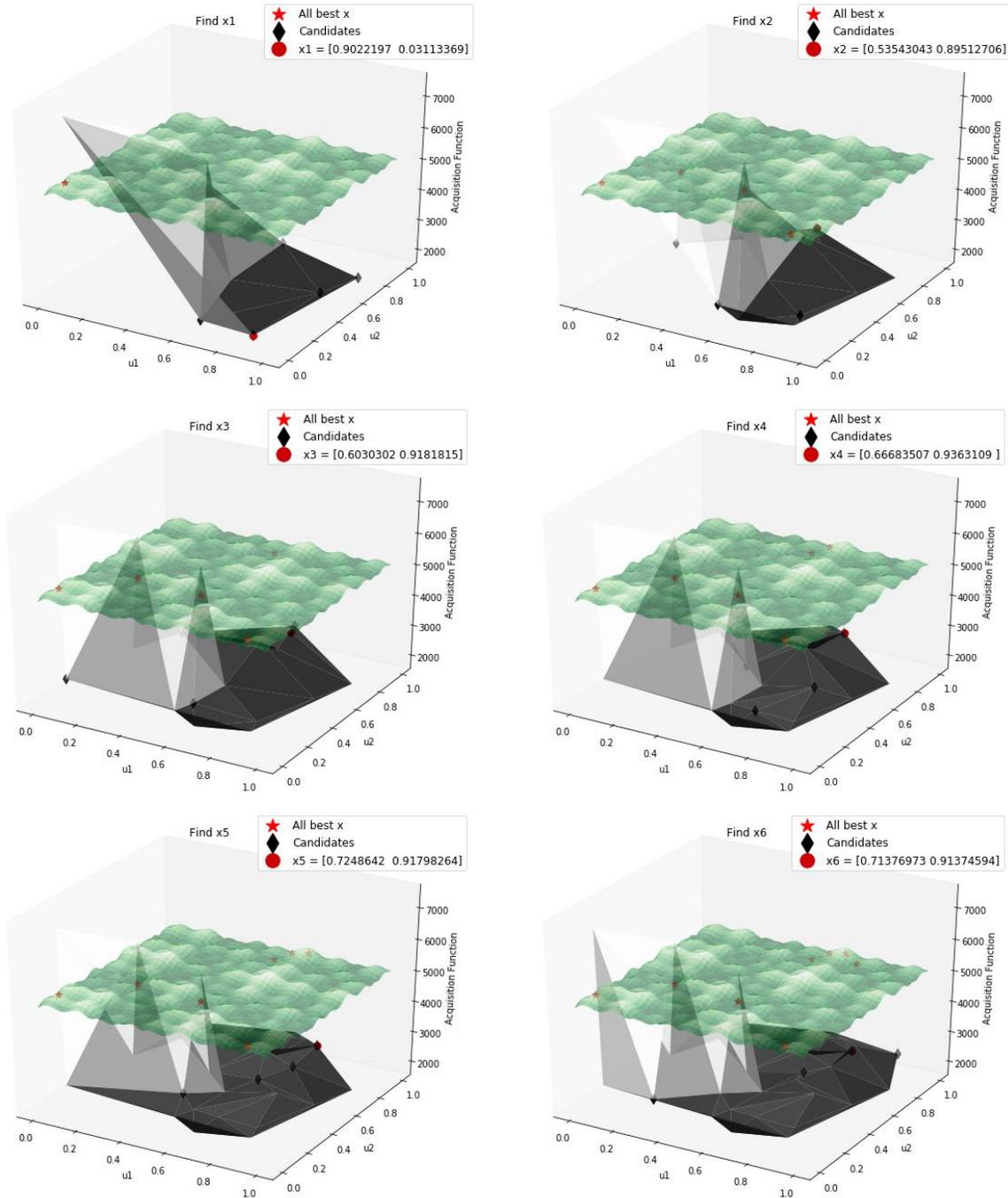

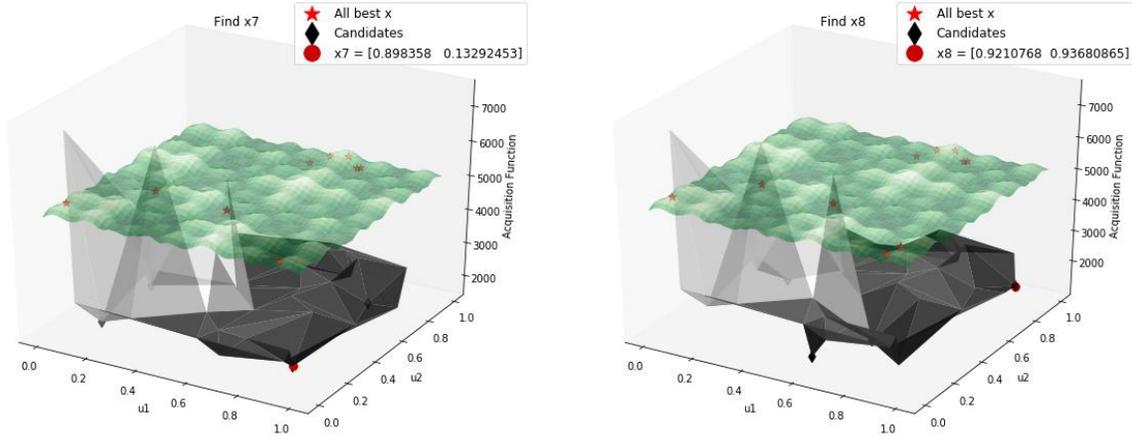

Fig. 5. 3-dimensional plotting for the sampling point position chosen during some iterations

### 5.3 Benchmarking using synthetic test functions

In this part, we benchmarked three widely used, non-strictly convex, single-objected, and lower-dimensional functions that are particularly hard to optimize due to the presence of multiple local or global minima or deep valley-like regions (Eggholder function, Rosenbrock function, McCormick function). Thus, they are well-known test problems for global optimization and have been widely used as the benchmarks for various optimization characteristics [34]. Fig. 6, Fig. 7, and Fig. 8 present the benchmark results by comparing the IBO algorithm to simplicial homology global optimization [35], dual annealing optimization [36], differential evolution [37], and basin-hopping [38], on three qualitatively different test functions mentioned above.

The red dot in the three figures represents the global minimum for given test functions. In Fig. 6, the global minimum for the Eggholder function locates in position (512, 404.2319, −959.6407). The IBO algorithm catches the solution (511.95715, 404.26605, −959.48926) shown as a black star sign in the figure, which performs exceptionally well comparing to other algorithms. The solutions generated by simplicial homology global optimization and dual annealing optimization are closer to the global minimum. Differential evolution and basin-hopping algorithms do not perform well on the Eggholder function. Their final solutions are far off from the global minimum.

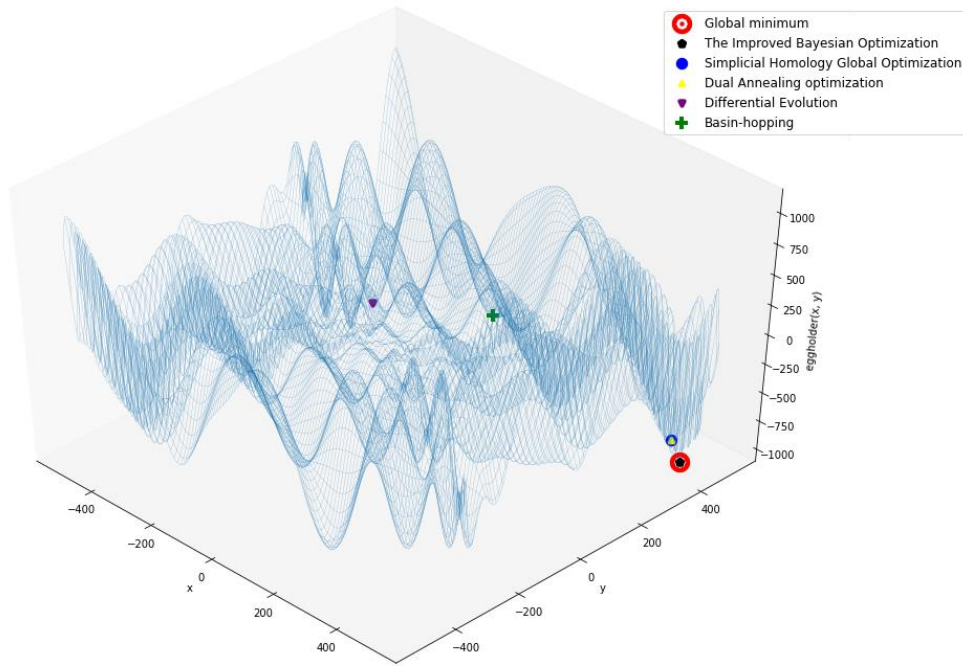

Fig. 6. Algorithms comparison for optimization of Eggholder function

Fig. 7 presents the computational results using the Rosenbrock function. The global minimum of 2D Rosenbrock is in (1, 1, 0). All five optimization algorithms perform well for reaching the global minimum within a reasonable amount of run time. Fig. 8 illustrates the computational results using the McCormick function. The global minimum for the McCormick function is located at (−0.54719, −1.54719, −1.9133). Our computational experiments show, IBO algorithm, dual annealing, and differential evolution outperform simplicial homology global optimization and basin-hopping. The IBO algorithm, dual annealing optimization, and differential evolution found the global minimum for the McCormick function, while both simplicial homology (blue dot) and the Basin-hopping algorithm ended (green cross) far from the global minimum.

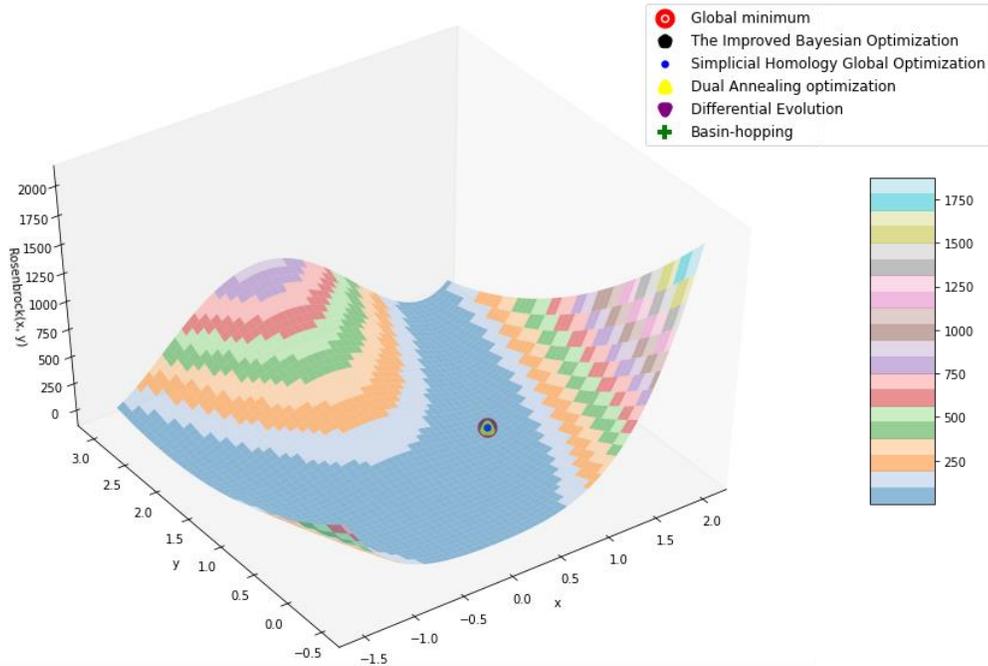

Fig. 7. Algorithms comparison for optimization of Rosenbrock function

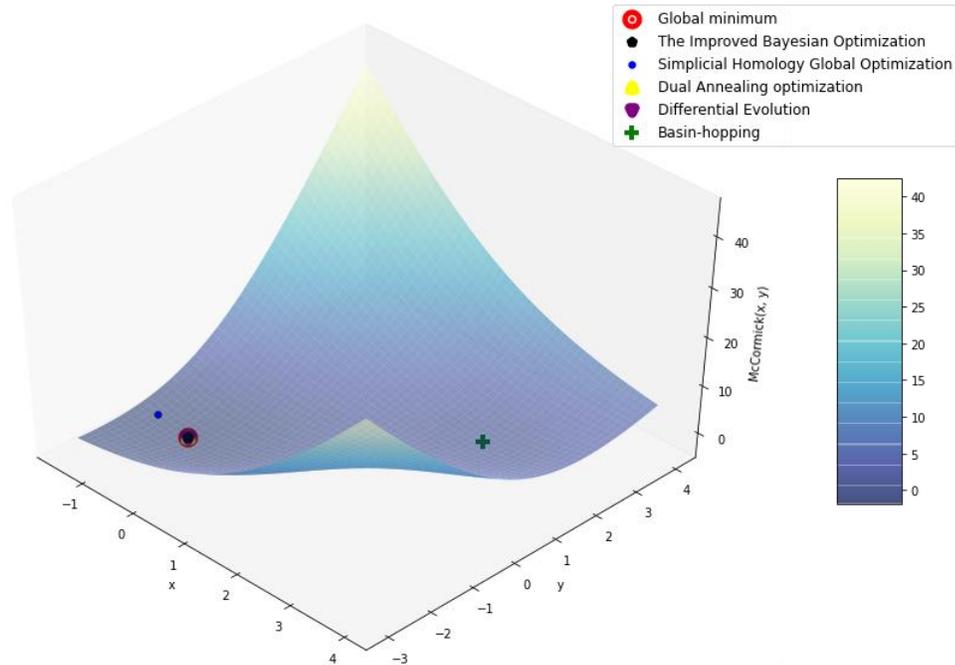

Fig. 8. Algorithms comparison for optimization of McCormick function

Based on our simulation experiments, the IBO algorithm for these objective functions shows the best performance. Compared to the other four optimization algorithms, the IBO algorithm performs well on the low-dimensional global optimization problems. It can reach a very accurate optimal solution for the three standard test functions.

### 5.4 Solving complex high-dimensional optimal control problem using the IBO algorithm

In this section, computational studies on the IBO algorithm is conducted on an SEIR epidemic model with a control variable defined in equations (3) – (8). Under investigation, the SEIR model has a control variable with 100 time-periods, so the problem's dimensionality is 100. The objective function of the optimal control problem is:

$$\text{Min } V = \int_0^{t_f} C_1 I(t) + C_2 \left| 0.3 \sin(10u(t)) + \sin(13u(t)) + 0.9 \sin(42u(t)) + 0.2 \sin(12u(t)) + u^2(t) \right|$$

The parameters $C_1$ and $C_2$ represent the cost of null control and the cost of control per individual in each time-period, respectively. The total cost of a control strategy is the sum of the cost at each time-period. For a control strategy with 100 time periods, it is necessary to calculate all state variables values at each time-period using the SEIR model in equations (4) – (8). Then it needs to calculate the cost using this complicated objective function 100 times for one control strategy. Therefore, computationally calculate the total cost is time-consuming. Fig. 9 shows a 2-dimensional plotting of the relationship between the control variable and the objective function. In Fig. 9, the control variable only has one time-period, and the objective function has multiple local minima. Doubtlessly, the objective function in the studied SEIR model with 100 time periods is more complicated.

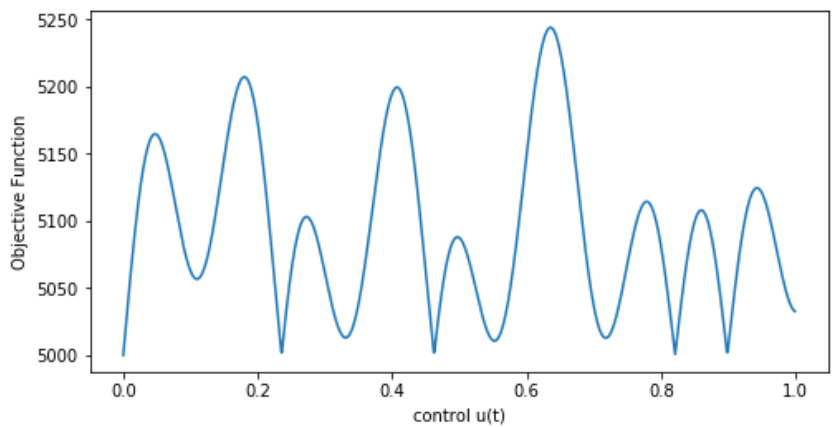

Fig. 9. 2-dimensional relationship plotting for control strategy with one time-period and objective function

Fig. 10 shows the comparison results of the SEIR epidemic model with control and without any control. The black and red dotted lines represent the trend for exposed and infected populations without control over time. The black and red solid lines represent the trend for exposed and infected populations with optimal control generated by the IBO algorithm. From Fig. 10, we can find that the infected population raised sharply initially when the model is without control and declined very slowly to zero. However, the optimal control strategy created by the IBO algorithm effectively and quickly control the epidemic once it starts to break out.

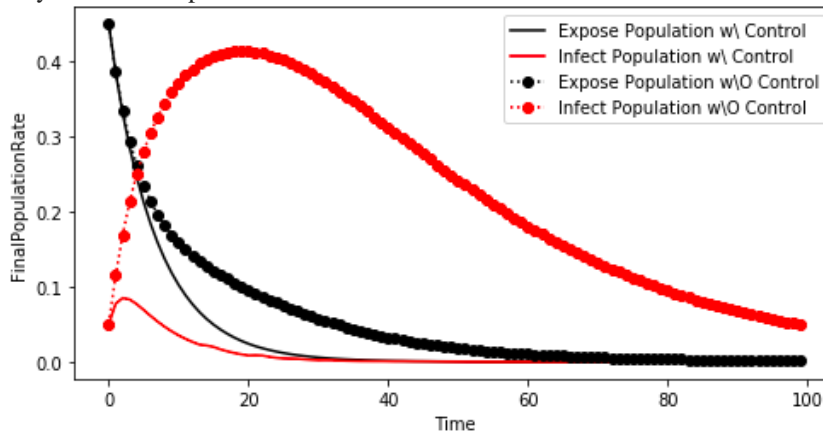

Fig. 10. The comparison of population rate for the IBO algorithm and null control

Next, the position updates of the consecutive sampling points are tracked from iteration to iteration. In our simulation experiments, we set the value of $l = 1$ in **Algorithm 1**, which means the IBO algorithm only retains one best candidate from all sampling candidates to be the next sampling point. The convergence plots in Figs. 11 and 12 depict how the distance changes between consecutive sampling points and how many iterations are required to find the global optima. The total iterations are the sum of the iterations $m$ of sampling the new next point and the iterations of local search. From Fig. 11, we can see that in the first 15 iterations, the distances between consecutive sampling points are relatively large and jumpy. It indicates that the sampling point is explored and updated by the global search, which was implemented from lines 5 to 17 in **Algorithm 1**. After that, the IBO algorithm starts to carry out more local searches within the while loop. Thus, it is easy to see the gradual downward trend in the later iterations. Fig. 12 shows the corresponding objective function values for the best sampling point at each iteration. Since we select the next sampling point according to the acquisition function value rather than the objective function value, that is why the trend line does not decrease monotonously within the first 15 iterations. After the local search process started, the IBO algorithm starts to perform searches based on the objective function value. Therefore, the trend line descends monotonously after the $15^{th}$ iteration. From both Figs. 11 and 12, one can observe a good convergence performance of the IBO algorithm. It almost explores the global optima around 30 iterations with the solution time is around 70 seconds. These computational studies indicate that the IBO algorithm works well and effective for especially high-dimensional search spaces.

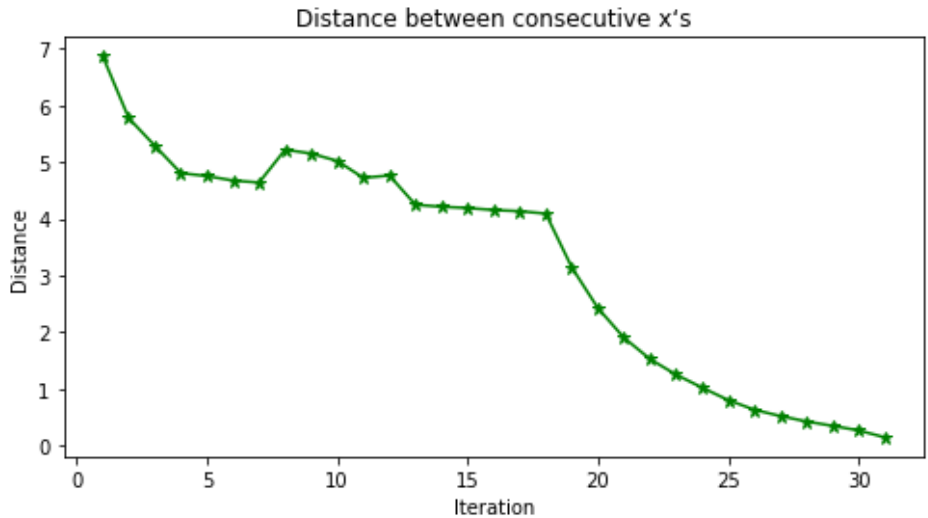

Fig. 11. The distance change between consecutive sampling points during iterations

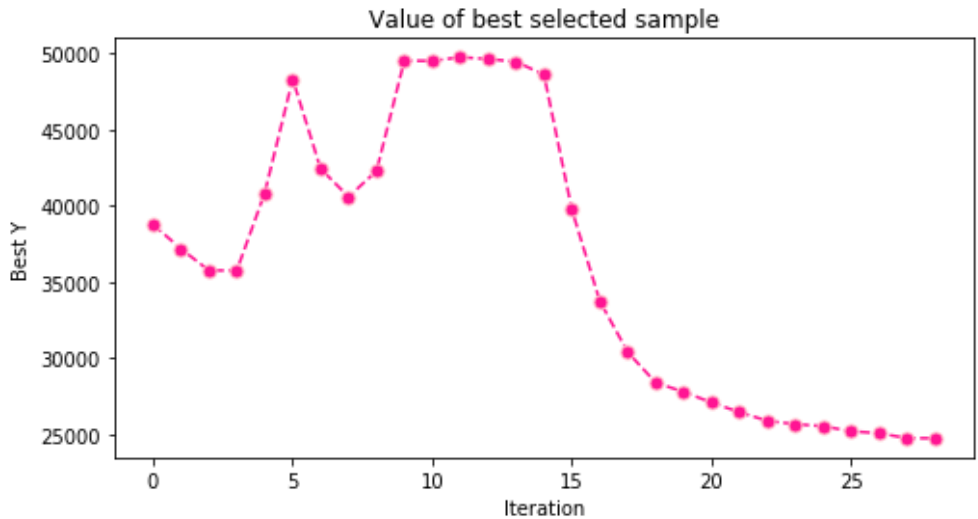

Fig. 12. The objective function value change between consecutive sampling points during iterations

## 5.5 Comparisons of the IBO algorithm, Random Search, PSO algorithm, and standard BO

For evaluating the accuracy performance of optimization, we compared our IBO algorithm with the standard BO algorithm, and two popular evolutionary algorithms (Random Search algorithm and PSO algorithm) [22]. To compare four algorithms under the same implementation condition, we do compute their mean squared error (MSE) values [39-40] on the same dataset named diabetes *datasets* from Python *sklearn* library [41]. The diabetes *database* contains sample data and corresponding actual target values [41]. Three algorithms need to predict the target values regarding sample data. Then calculate the MSE values between prediction target values and real target values. Since all evaluation criteria follow the convention that higher return values are better than lower return values [42]. We compared the MSE of the three algorithms here, and the MSE values are positive. It means that the algorithm performs better with smaller MSE values. Thus, to follow the evaluation criteria' higher is better', we used the negative MSE values for three algorithms. We choose the existing function 'neg_mean_squared_error' in Python [42] to calculate each algorithm's negative MSE values. The simulation results on the same diabetes *datasets* are shown below:

Random Search neg. MSE = −3511.23
PSO algorithm neg. MSE = −3380.56
Standard BO algorithm neg. MSR = −3367.93
The IBO algorithm neg. MSE = −3146.28

Fig. 13. plots the negative MSE values of three algorithms in each iteration. The results and figure show the IBO algorithm starts to return a higher score within a smaller number of iterations than Random Search algorithm, PSO algorithm and standard BO algorithm.

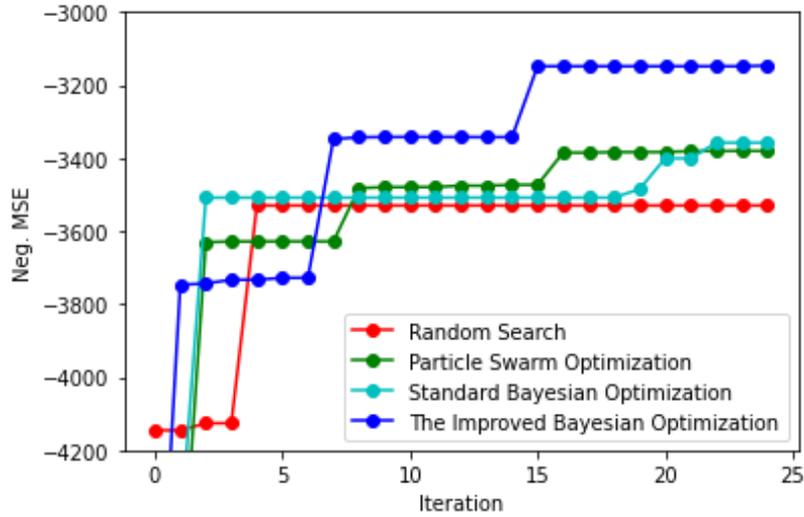

Fig. 13. The negative MSE comparison of four algorithms

Besides, we conducted simulation experiments to exam the robustness of the IBO algorithm, Random Search, and PSO algorithm with the different initial solutions, and the results are summarized in Fig. 14. In Fig. 14, the horizontal axis represents the initial control input of the algorithm that each dimension is set up the constant $\varphi \in [u_l, u_u]$ that is $u^0 = \left(u_0^0 = \varphi, \ldots, u_{t_f}^0 = \varphi\right)$. The results show that the PSO algorithm is more sensitive to the initial solutions than the IBO algorithm and the Random Search. Besides, Random Search and PSO algorithms have a poor performance to reach the global optima. Although Random Search behaves relatively stable, it is trapped in some local optima, so their objective function values are larger than the objective function value corresponding to the global optima. While the IBO algorithm performs relatively stable with different initial solutions, it consistently searches out the global minimum for the given optimal control model.

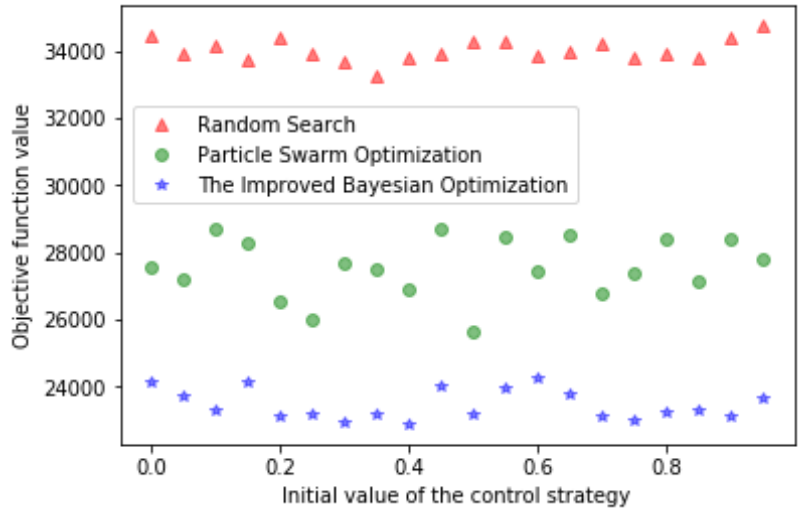

Fig. 14. The sensitivity analysis of different initial control input for three algorithms

## 6. Conclusion

In this paper, we propose an Improved Bayesian Optimization (IBO) approach by considering more suboptimal steps for optimizing the acquisition function process. An Adam-based local optimization process is used to improve the current solution even further when the best candidate is generated from acquisition function optimizations. Different kernel functions and acquisition function choices are discussed to study their impacts on the IBO algorithm's performances. We extend the IBO algorithm to solve both low-dimensional optimization problems with multiple local minima and high-dimensional dynamic optimal control models. Computational benchmarks of the IBO algorithm are also compared against four existing optimization algorithms, namely Simplicial homology global optimization, Dual annealing optimization, Differential evolution, and Basin-hopping algorithm, on three widely used single-objective test functions. Our benchmark tests showed that the IBO algorithm could reach the most precise solutions for all three test functions. The IBO algorithm performs an excellent optimization characteristic in low-dimensional optimization problems. Also, the trajectory of distances between consecutive sampling points indicates that implementing a local optimization process after the acquisition function optimizations is useful and necessary. Besides, compared with the Random Search and PSO algorithm, the experimental results demonstrate that the IBO algorithm can achieve better optimization results for the high-dimensional dynamic optimal control model. The IBO algorithm is also more robust with different initial control strategies than Random Search and PSO algorithm. In Bayesian inference, probability represents the degree of belief. It's possible to accurately catch a better candidate with high posterior belief as the next sampling point if the algorithm well utilizes the Bayesian inference. Thus, one possible research direction would be the extension of our IBO algorithm based on Bayesian inference.